\documentclass[aps,prc,twocolumn,superscriptaddress,showpacs]{revtex4}

\usepackage{graphicx}
\usepackage{dcolumn}
\usepackage{bm}
\usepackage{ulem}
\usepackage{color}

\newcommand{\al}{$\alpha$}

\newcommand{\raa}{($\alpha$,$\alpha$)}

\newcommand{\rag}{($\alpha$,$\gamma$)}
\newcommand{\ran}{($\alpha$,n)}

\newcommand{\rna}{(n,$\alpha$)}
\newcommand{\stot}{$\sigma_{\rm{reac}}$}
\newcommand{\sred}{$\sigma_{\rm{red}}$}
\newcommand{\ered}{$E_{\rm{red}}$}
\newcommand{\cer}{$^{140}$Ce}
\newcommand{\pras}{$^{141}$Pr}

\newcommand{\pmiv}{$^{144}$Pm}

\newcommand{\cet}{$^{140}$Ce}

\begin{document}

\title{
  Total reaction cross section
  $\sigma_{\rm{reac}}$ of $\alpha$-induced reactions from elastic scattering:
  the example $^{140}$Ce($\alpha$,$\alpha$)$^{140}$Ce
}

\author{Peter Mohr}
\email{WidmaierMohr@t-online.de}
\affiliation{
Diakonie-Klinikum, D-74523 Schw\"abisch Hall, Germany}
\affiliation{
Institute of Nuclear Research (ATOMKI), H-4001 Debrecen, Hungary}

\date{\today}

\begin{abstract}
Angular distributions of elastic $^{140}$Ce($\alpha$,$\alpha$)$^{140}$Ce
scattering are analyzed in the framework of the optical model from low
energies around the Coulomb barrier up to about 40\,MeV. 
From the local fits the total reaction cross
section $\sigma_{\rm{reac}}$ is extracted. This procedure requires
experimental scattering data with small uncertainties. The results for
$\sigma_{\rm{reac}}$ are compared to predictions of different systematic
global potentials. It is found that the total reaction cross section
$\sigma_{\rm{reac}}$ is well predicted from all global potentials under study
although the reproduction of the angular distributions is not perfect in all
cases. In addition, the lower energy limit for the extraction of
$\sigma_{\rm{reac}}$ from elastic scattering angular distributions is
analyzed. Finally, the potentials under study are used to calculate the
$^{143}$Nd(n,$\alpha$)$^{140}$Ce cross section, and the predictions are
compared to experimental data.
\end{abstract}


\maketitle

\section{Introduction}
\label{sec:intro}
The total reaction cross section \stot\ is a basic quantity for the prediction
of reaction cross sections in the statistical model. It has been found that
especially at low energies huge discrepancies are found between predictions
for \al -induced reaction cross sections using different global \al -nucleus
optical potentials. This holds in particular for \rag\ capture reactions for
targets with masses above $A \approx 100$
\cite{Som98,Gyu06,Ozk07,Cat08,Yal09,Gyu10,Kis11,Kis12}, but also the energy
dependence of recent \ran\ data for \pras\ was difficult to fit
\cite{Sau11,Mohr11}.

The total reaction cross section \stot\ is related to the complex scattering
matrix $S_L = \eta_L \, \exp{(2i\delta_L)}$ by the well-known relation
\begin{equation}
\sigma_{\rm{reac}} = \sum_L \sigma_L 
   = \frac{\pi}{k^2} \sum_L (2L+1) \, (1 - \eta_L^2)
\label{eq:stot}
\end{equation}
where $k = \sqrt{2 \mu E_{\rm{c.m.}}}/\hbar$ is the wave number,
$E_{\rm{c.m.}}$ is the energy in the center-of-mass (c.m.)\ system, and
$\eta_L$ and $\delta_L$ are the real reflexion coefficients and scattering
phase shifts. $\sigma_L$ is the contribution of the $L$-th partial wave to the
total reaction cross section \stot . The relation in Eq.~(\ref{eq:stot}) has
recently been confirmed experimentally at low energies \cite{Gyu12}.

The present study determines \stot\ from
the angular distributions at $E_{\rm{lab}} = 14.7$\,MeV \cite{Wat71} and 19.0,
24.0, 32.0, and 37.7\,MeV \cite{Gua94}. The latter data are unfortunately not
published. The data have been measured at the XTU Tandem of the INFN
Laboratori Nazionali di Legnaro using a cerium target enriched to about 96\,\%
in the semi-magic $N=82$ nucleus \cer . Access to the numerical \cer \raa
\cer\ scattering data of \cite{Gua94} will be provided via the EXFOR database
\cite{EXFOR}. In addition, elastic scattering data for natural cerium (\cer :
88.48\,\%) are available in literature. These data have also been analyzed,
and the results fit into the observed systematics of total reaction cross
sections. However, the angular distributions are not shown here because they
may be affected by the other cerium isotopes.

The obtained results for \stot\ are compared to predictions from
recently published \al -nucleus potentials \cite{Avr09,Avr10,Mohr12} and to
the widely used potential by McFadden and Satchler \cite{McF66}.
The example of \cer\ is well-suited for a study of uncertainties for \stot\ at
low energies because experimental scattering data are available down to
relatively low energies. Further information on the \al -nucleus potential can
be derived from reaction data at very low energies. Often the
\ran\ reaction has been used for this purpose (see e.g.\ \cite{Sau11,Mohr11}),
but in the present case there are no experimental data for the \cer \ran
$^{143}$Nd reaction. Instead, the reverse $^{143}$Nd\rna \cer\ reaction will
be investigated here.

\section{Optical model analysis}
\label{sec:OM}
The elastic scattering angular distributions are analyzed in the usual way
within the framework of the optical model and a complex \al -nucleus optical
potential. Details of the fitting procedure for this local fit are similar to
Refs.~\cite{Mohr11,Mohr12} and are only briefly summarized here. 

The total potential $U(r)$ is composed of the real Coulomb potential $V_C(r)$
and the complex nuclear potential $V(r) + iW(r)$. The Coulomb potential is
calculated from a homogenously 
charged sphere where the Coulomb radius is taken from the root-mean-square
($r.m.s.$) radius of the real folding potential (see below). Properties of the
real and imaginary part of the nuclear potential are presented in the next
paragraphs.

\subsection{Real part of the nuclear potential}
\label{sec:real}
The real
part of the nuclear potential is determined by a double-folding procedure
where the folding potential $V_F(r)$ is modified by a strength parameter
$\lambda$ and a width parameter $w$: 
\begin{equation}
V(r) = \lambda \, V_F(r/w) \quad .
\end{equation}
The strength parameter $\lambda$ and the width parameter $w$ will be adjusted
to the experimental \cet \raa \cet\ angular distributions. Obviously, the
width parameter $w$ should remain close to unity; otherwise, the folding
potential would be questionable. The strength parameter $\lambda$ is typically
around 1.1 to 1.4 leading to volume integrals per interacting nucleon pair of
$J_R \approx 310 - 350$\,MeV\,fm$^3$ \cite{Atz96}. (As usual, the
negative signs of $J_R$ and $J_I$ are neglected in the following discussion.)

The calculation of the folding potential requires the density of the
\cer\ target 
nucleus which is usually derived from electron scattering. Although no data
are available in the widely used compilation \cite{Vri87},
two papers \cite{Mil88,Kim92} use a \cer\ density and refer to a 
``complementary, high-precision, elastic electron scattering experiment
performed at Saclay'' with the reference ``D.\ Goutte {\it et al.}, to be
published''. The density parameters in Fourier-Bessel parametrization can be
found in an underlying Ph.D.\ thesis by B.\ L.\ Miller \cite{Mil94}, Table
A.1, and are repeated here for easier access in future work (see Table
\ref{tab:ce140fb}). The resulting folding potential is very close to the
average of the potentials for the neighboring $N=82$ nuclei $^{139}$La and
$^{142}$Nd; thus, the reliability of the unpublished electron density of
\cer\ from \cite{Mil94} is confirmed.
\begin{table}[tbh]
\caption{\label{tab:ce140fb}
Experimental charge density distribution of \cet\ as listed in Table A.1 of
\cite{Mil94}, parametrized by a Fourier-Bessel series: $\rho(r) = \sum_n G_n
\times j_0(n \pi r/R_{FB})$ with the Fourier-Bessel cutoff radius $R_{FB} =
10$\,fm. 
}
\begin{center}
\begin{tabular}{rrcrr}
\multicolumn{1}{c}{$n$}
& \multicolumn{1}{c}{$G_n$}
&
& \multicolumn{1}{c}{$n$}
& \multicolumn{1}{c}{$G_n$} \\
\hline
1 & 0.759956   & &  2 & 0.578531 \\
3 & -0.589376  & &  4 & -0.184926 \\
5 & 0.337939   & &  6 & 0.006674 \\
7 & -0.132665  & &  8 & 0.002477 \\
9 & 0.002589   & & 10 & -0.013456 \\
\hline
\end{tabular}
\end{center}
\end{table}

\subsection{Imaginary part of the nuclear potential}
\label{sec:imag}
The imaginary part of the optical potential is parametrized by the usual
Woods-Saxon potentials of volume and surface type. Except at the highest
energy under study where an additional volume Woods-Saxon potential is
required, only a surface Woods-Saxon potential was used:
\begin{equation}
W(r) = W_V f(x_V) + W_S \frac{df(x_S)}{dx_S} \quad .
\label{eq:WS}
\end{equation}
$W_i$ are the depth parameters of the volume and surface imaginary potential,
and the Woods-Saxon function $f(x_i)$ is given by
\begin{equation}
f(x_i) = \Bigl[ 1 + \exp{(x_i)} \Bigr]^{-1}
\label{eq:WSshape}
\end{equation}
with $x_i = (r - R_i A_T^{1/3})/a_i$ and $i = V,S$ for the volume and surface
part. Note that $W_V < 0$ and $W_S > 0$ in the chosen conventions
(\ref{eq:WS}) and (\ref{eq:WSshape}) for an absorptive negative $W(r) < 0$. 
The maximum depth of the surface imaginary potential is given by $-W_S/4$ at
$r = R_S A_T^{1/3}$.

\subsection{Results}
\label{sec:res}
\begin{figure}[htb]
\includegraphics[bbllx=40,bblly=20,bburx=475,bbury=695,width=\columnwidth,clip=]{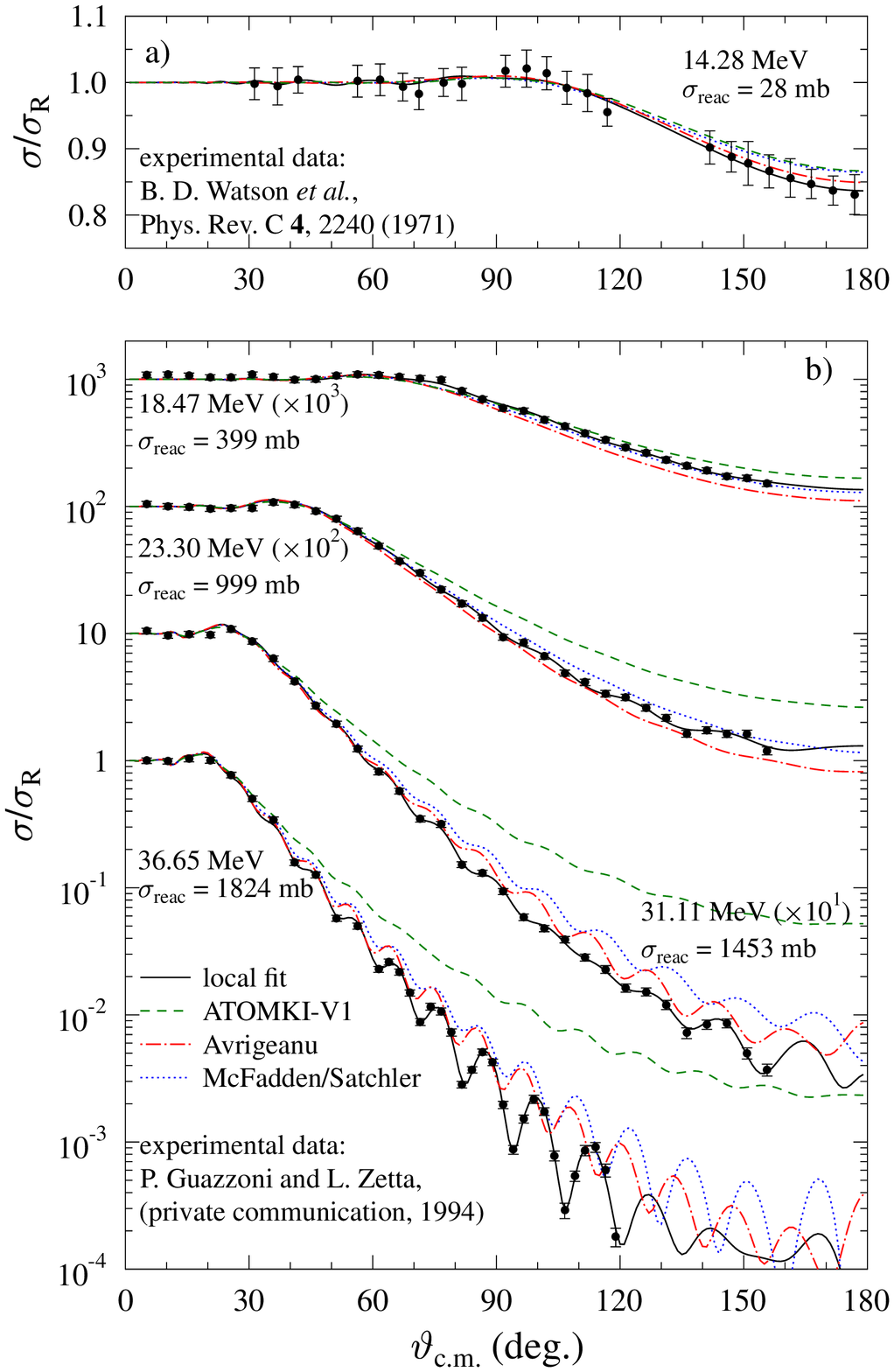}
\caption{
\label{fig:scat}
(Color online)
Rutherford normalized elastic scattering cross sections of the \cer \raa
\cer\ reaction versus the angle in center-of-mass frame. The lines are
calculated from a local potential fit which is adjusted to the scattering data
(full black line) and from different global \al -nucleus potentials 
\cite{McF66,Avr10,Mohr12} (MCF: blue dotted; AVR: red dash-dotted;
ATOMKI-V1: green dashed). The experimental data are taken from
\cite{Wat71,Gua94}. The given energies are $E_{\rm{c.m.}}$ in the
center-of-mass system.
}
\end{figure}

The parameters of the nuclear potential are adjusted to the experimental
angular distributions using a standard $\chi^2$ minimizing procedure.
The resulting parameters are listed in Table \ref{tab:pot}, and the fits are
compared to the experimental angular distributions in
Fig.~\ref{fig:scat}. Excellent agreement between the local fits and the
experimental data is obtained at all energies under study. For the real part
the parameters of the fits show very minor variations with energy and are in
their expected ranges \cite{Atz96}. The parameters of the real and imaginary
parts of the potential will be discussed in detail below.
\begin{table*}[tbh]
\caption{\label{tab:pot}
Parameters of the optical potential and the total reaction
cross section \stot\ derived from \cet \raa \cet\ angular
distributions.
}
\begin{center}
\begin{tabular}{cccccccccrcccccr@{$\pm$}lc}
\multicolumn{1}{c}{$E_{\rm{lab}}$} 
& \multicolumn{1}{c}{$E_{\rm{c.m.}}$}
& \multicolumn{1}{c}{$\lambda$}
& \multicolumn{1}{c}{$w$}
& \multicolumn{1}{c}{$J_R$}
& \multicolumn{1}{c}{$r_{R,{\rm{rms}}}$}
& \multicolumn{1}{c}{$W_V$}
& \multicolumn{1}{c}{$R_V$}
& \multicolumn{1}{c}{$a_V$}
& \multicolumn{1}{c}{$W_S$}
& \multicolumn{1}{c}{$R_S$}
& \multicolumn{1}{c}{$a_S$}
& \multicolumn{1}{c}{$J_I$}
& \multicolumn{1}{c}{$r_{I,{\rm{rms}}}$}
& \multicolumn{1}{c}{$\chi^2/F$}
& \multicolumn{2}{c}{\stot\ \footnote{from the local potential fit using
    Eq.~(\ref{eq:stot}); uncertainties estimated from the model-independent
    phase shift analysis}} 
& \multicolumn{1}{c}{Ref.} \\
\multicolumn{1}{c}{(MeV)} 
& \multicolumn{1}{c}{(MeV)}
& \multicolumn{1}{c}{(--)}
& \multicolumn{1}{c}{(--)}
& \multicolumn{1}{c}{(MeV\,fm$^3$)}
& \multicolumn{1}{c}{(fm)}
& \multicolumn{1}{c}{(MeV)}
& \multicolumn{1}{c}{(fm)}
& \multicolumn{1}{c}{(fm)}
& \multicolumn{1}{c}{(MeV)}
& \multicolumn{1}{c}{(fm)}
& \multicolumn{1}{c}{(fm)}
& \multicolumn{1}{c}{(MeV\,fm$^3$)}
& \multicolumn{1}{c}{(fm)}
& \multicolumn{1}{c}{(--)}
& \multicolumn{2}{c}{(mb)} 
& \multicolumn{1}{c}{Exp.} \\
\hline
14.69 & 14.28 & 1.146 & 1.021  & 315.8 & 5.641 
& \multicolumn{3}{c}{--} & 28.0 & 1.444 & 0.507 & 18.2 & 7.774 
& 0.1 & 28  & 4~ & \cite{Wat71} \\ 
19.0 & 18.47 & 1.218 & 1.014 & 328.1 & 5.599 
& \multicolumn{3}{c}{--} & 101.8 & 1.517 & 0.303 & 43.1 & 7.970 
& 0.7 & 399 & 30~ & \cite{Gua94} \\ 
23.97 & 23.30 & 1.253 & 1.017 & 340.7 & 5.616 
& \multicolumn{3}{c}{--} & 114.0 & 1.445 & 0.434 & 63.2 & 7.709 
& 0.8 & 999 & 30~ & \cite{Gua94} \\ 
32.0 & 31.11 & 1.276 & 1.006 & 336.0 & 5.557 
& -21.1 & 1.311 & 0.211 & 13.4 & 1.574 & 0.416 & 58.6 & 5.860 
& 0.8 & 1453 & 44~ & \cite{Gua94} \\
37.7 & 36.65 & 1.265 & 0.999 & 326.3 & 5.518 
& -19.1 & 1.374 & 0.269 &  8.9 & 1.580 & 0.729 & 62.6 & 6.218 
& 1.0 & 1824 & 55~ & \cite{Gua94} \\
45.0\footnote{natural cerium (\cer : 88.48\,\%)} 
     & 43.75 & 1.195 & 1.020 & 328.3 & 5.635 
& -28.2 & 1.028 & 0.809 & 28.0  & 1.516 & 0.504 & 59.3 & 6.291
& 0.4 & 1940 & 97\footnote{increased 5\,\% uncertainty because of
  contributions from other cerium isotopes} & \cite{Bak72} \\
\hline
\end{tabular}
\end{center}
\end{table*}

The width parameter $w$ is about 1.5\,\% larger than unity and practically
constant except at the highest energy for \cer . 
The strength parameter $\lambda$ varies by about 10\,\% between 1.15 and 1.26;
this leads to a similar variation of the real volume integral $J_R$ which
shows a maximum of 341\,MeV\,fm$^3$ at 24\,MeV and slightly smaller values at
higher and lower energies (similar to the finding in \cite{Mohr00}). The
variation of the potential parameters of the real part is relatively small,
and thus the real part of the potential and its energy dependence should not
be the reason for major uncertainties in the prediction of \al -induced
reaction cross sections at low energies.

Contrary to the real part, the imaginary part shows a significant energy
dependence. The most obvious signature is the change from a surface
Woods-Saxon potential at very low energies to a dominating volume Woods-Saxon
potential at the higher energies under study. It is sufficient to use a
pure surface Woods-Saxon potential at the lower energies below 25\,MeV, and
reduced $\chi^2/F$ values around unity are obtained. The variation of the
geometry of the surface imaginary part at the lower energies is small: the
radius parameter $R_S$ varies by less than 5\,\% around its average value, and
the diffuseness $a_S$ shows a somewhat larger spread of about 25\,\% around
its average. However, there is no systematic energy dependence of $R_S$ and
$a_S$ which may lead to uncertainties in the extrapolation of the potential
down to very low energies (see also Sect.~\ref{sec:sens}). The depth parameter
$W_S$ and the imaginary volume integral $J_I$ increase with energy for the
lowest 3 energies.

\begin{figure}[htb]
\includegraphics[bbllx=30,bblly=20,bburx=475,bbury=445,width=\columnwidth,clip=]{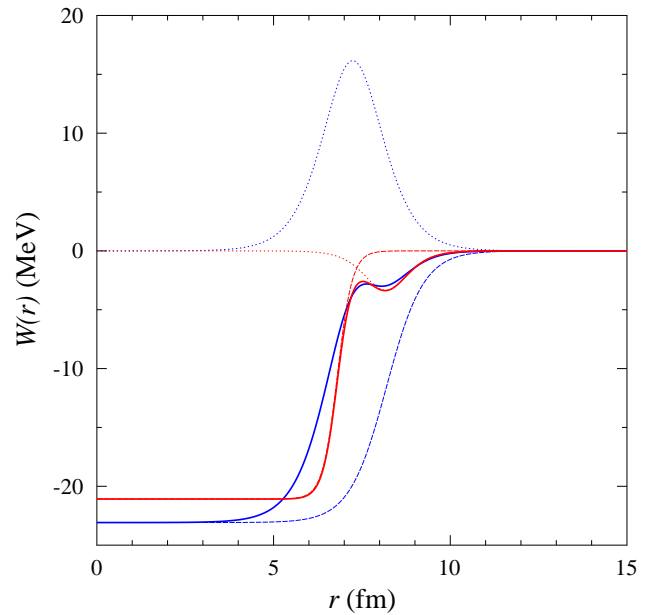}
\caption{
\label{fig:pot_imagamb}
(Color online)
Ambiguity of the imaginary potential, illustrated for the analysis of the
angular distribution at $E = 31.11$\,MeV. The imaginary potentials of two
different fits (shown in blue and red full lines) are very similar. However,
the decomposition into the volume (thin dashed) and surface (thin dotted)
components shows that in one case there is a weak negative surface term (red)
whereas in the other case a strong positive surface term (blue) is found.
}
\end{figure}

At higher energies above 30\,MeV, the best $\chi^2/F$ is obtained from a
combination of a volume and a surface Woods-Saxon potential. At all energies
above 30\,MeV the volume term is dominating. An interesting ambiguity in the
parametrization of the imaginary potential is found for the combination of
volume and surface Woods-Saxon poentials. Very similar potentials can be
obtained using a positive or a negative surface term. This is illustrated in
Fig.~\ref{fig:pot_imagamb} for the analysis at $E = 31.11$\,MeV. In the first
fit (parameters given in Table 
\ref{tab:pot}) the usual behavior is found: the weak negative surface
component modifies the dominating volume component in the surface region
around 8\,fm. However, in the second fit (parameters: $\lambda = 1.269$, $w =
1.004$, $W_V = -23.1$\,MeV, $R_V = 1.579$\,fm, $a_V = 0.525$\,fm, $W_S =
-64.6$\,MeV, $R_S = 1.394$\,fm, $a_S = 0.571$\,fm, \stot\ = 1464\,mb) a
relatively strong and positive surface contribution is found. But the total
imaginary part is very similar in both fits, in particular in the most
relevant surface region, and thus also the resulting
$\chi^2/F$ and derived \stot\ are practically identical. 
Such an ambiguity is found at
all energies above 30\,MeV. In principle, the fit with the lowest $\chi^2/F$
should be used for the extraction of \stot\ from the experimental angular
distribution. However, because the found $\chi^2/F$ minima are very similar
and do not differ by more than 10\,\%, and because the derived \stot\ are
stable within 1\,\%, the fits with the usual negative surface component are
listed in Table \ref{tab:pot}. This choice should avoid further
complications e.g.\ when using these parameters for the construction of a
global potential. For completeness it is pointed out that fits with
a pure volume Woods-Saxon imaginary potential have a worse
$\chi^2/F$, and the derived \stot\ vary within several per cent. 

It is obvious that the energy dependence of $J_I$ and the shape of the
imaginary part of the potential are dominating sources for
uncertainties of cross section predictions at very low energies. As usual, the
imaginary potential $W(r)$ is dominated by the surface contribution at low
energies, and its volume integral $J_I$ increases with energy. At higher
energies above 30\,MeV the volume contribution becomes dominant. The data at
$23.30$\,MeV could also be fitted using a combination of volume ($W_V =
-13.8$\,MeV, $R_V = 1.30$\,fm, $a_V = 0.81$\,fm) and surface ($W_S =
13.6$\,MeV, $R_S = 1.58$\,fm, $a_S = 0.34$\,fm) potential, i.e.\ with a further
decreasing volume contribution. Although the number of ajustable parameters is
larger, this fit did not lead to an improved description of the angular
distribution (no significant improvement of $\chi^2/F$). Therefore, from this
energy down to lowest energies only a surface imaginary potential was
used. This choice is confirmed by the fact that an excellent reproduction of
all angular distributions below 25\,MeV with $\chi^2/F \lesssim 1$ has been
achieved using a pure surface Woods-Saxon potential in the imaginary part. 
Because the volume integrals $J_I$ increase with energy at these low
energies, a similar behavior is found for the strength of the surface
imaginary potential $W_S$.

The influence of uncertainties in the real part and the imaginary part on the
calculation of the total reaction cross section \stot\ will be investigated in
further detail later (see Sect.~\ref{sec:stot}).

\begin{figure}[htb]
\includegraphics[bbllx=30,bblly=20,bburx=475,bbury=800,width=\columnwidth,clip=]{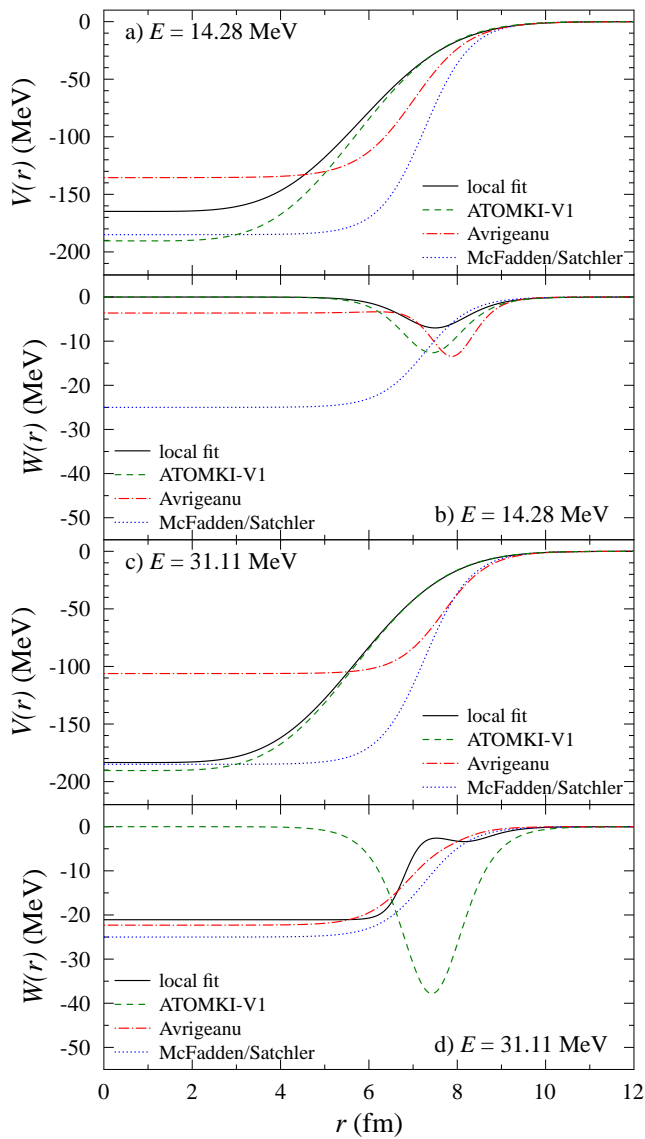}
\caption{
\label{fig:pot}
(Color online)
Real part (a,c) and imaginary part (b,d) of the underlying potentials for the
angular distributions at 14.28\,MeV (a,b) and 31.11\,MeV (c,d) from the local
fit (black full line) and the global potentials \cite{McF66,Avr10,Mohr12}
(MCF: blue dotted; AVR: red dash-dotted; ATOMKI-V1: green dashed; colors and
linestyles identical to Fig.~\ref{fig:scat}). A decomposition of the best-fit
potential at 31.11\,MeV into its volume and surface components is also given
(thin dashed and dotted black lines).
}
\end{figure}

The elastic scattering angular distributions are also compared to predictions
from several global \al -nucleus optical potentials. In Fig.~\ref{fig:scat} we
show the very simple 4-parameter potential by McFadden and Satchler (MCF)
\cite{McF66}, the many-parameter potential by Avrigeanu (AVR) \cite{Avr10},
and the recent first version of the ATOMKI potential (ATOMKI-V1) \cite{Mohr12}.
For illustration, I also show the real and imaginary nuclear potentials for
the lowest energy of 14.28\,MeV and an intermediate energy (31.11\,MeV) in
Fig.~\ref{fig:pot}. 

\section{Determination of the total reaction cross section \stot\ and the
  reduced cross section \sred }
\label{sec:stot}
\subsection{Extraction of \stot\ and \sred\ from the experimental angular
  distributions}
\label{sec:stot_exp}
The total reaction cross sections \stot\ are derived from these
local fits using Eq.~(\ref{eq:stot}).
For an estimation of the uncertainty of \stot\ an additional 
model-independent phase shift analysis (PSA) has been performed using the
technique of \cite{Chi96}. The total reaction cross sections
\stot\ from the PSA
are close to the results of the local potential fit. A typical
uncertainty for \stot\ of about 3\,\% is estimated (see also \cite{Mohr10})
except at very low energies significantly below the Coulomb barrier.

It is interesting to note that there is only a relatively weak dependence of
the derived total reaction cross section \stot\ on the absolute normalization
of the elastic scattering angular distribution. At relatively low energies the
absolute normalization is well-defined from Rutherford scattering at forward
angles. This is the case for all energies under study in this work.
But even at energies far above the Coulomb barrier it turns out that
the potentials are relatively well-defined by the diffraction pattern in the
angular distribution. Consequently the extracted \stot\ vary typically by
less than 10\,\% even for strong variations of the absolute normalization of
up to 30\,\%.

In general, the extraction of the total reaction cross section
\stot\ from elastic scattering angular distributions requires theoretical
considerations and is thus somewhat 
model-dependent. But because of the small sensitivity of \stot\ to
the chosen model, the total reaction cross section \stot\ can be considered as
a quasi-experimental quantity. This holds in particular for the present case
where the elastic scattering angular distributions cover a wide angular
range and the scattering energies are not extremely low. 
These quasi-experimental total reaction cross sections are compared to
predictions of global \al -nucleus potentials in Table \ref{tab:sigtot}. The
very simple MCF potential \cite{McF66},
the many-parameter AVR potential \cite{Avr10}, and the few-parameter
ATOMKI-V1 potential \cite{Mohr12} provide
almost identical 
\stot\ in good agreement with the experimental results although the
reproduction of the experimental angular distributions is not perfect in 
some cases (see Fig.~\ref{fig:scat}). Obviously, the global potentials (which
are not fitted to the data) cannot reproduce the experimental angular
distributions with the same quality as the local fit. In particular, the
ATOMKI-V1 potential has been adjusted to low-energy scattering data only.
This leads to a relatively poor description of the experimental angular
distributions at higher energies because of the missing volume term in the
imaginary part. However, even deviations in the angular distribution of up to
one order of magnitude at very backward angles do not prevent the prediction
of \stot\ with minor uncertainties.
\begin{table}[htb]
\caption{\label{tab:sigtot}
Experimental total reaction cross sections \stot\ (in mb) derived from \cer
\raa \cer\ elastic scattering angular distributions (from Table
\ref{tab:pot}), compared to predictions from the global \al -nucleus
potentials MCF \cite{McF66}, AVR \cite{Avr10}, and ATOMKI-V1
\cite{Mohr12}. For a discussion of the $^{143}$Nd\rna \cer\ data see
Sect.~\ref{sec:na}.
}
\begin{center}
\begin{tabular}{cr@{$\pm$}lrrr}
\multicolumn{1}{c}{$E_{\rm{c.m.}}$ (MeV)}
& \multicolumn{2}{c}{exp.}
& \multicolumn{1}{c}{MCF}
& \multicolumn{1}{c}{AVR}
& \multicolumn{1}{c}{ATOMKI-V1} \\
\hline
10.0~\footnote{corresponds to the low-energy $^{143}$Nd\rna \cer\ data}
  & \multicolumn{2}{c}{$(1.9 \pm 0.4\,\mu{\rm{b}})$}\footnote{derived from the
low-energy $^{143}$Nd\rna \cer\ data}
  & $9.3\,\mu$b  & $0.7\,\mu$b  & $4.5\,\mu$b  \\
14.28~  &   28  &  4  &   26  &   23  &   24  \\
18.47~  &  399  & 30  &  464  &  477  &  460  \\
23.30~  &  999  & 30  &  987  & 1025  &  989  \\
31.11~  & 1453  & 44  & 1472  & 1508  & 1494  \\
36.65~  & 1824  & 55  & 1676  & 1712  & 1712  \\
43.75\footnote{natural cerium (\cer : 88.48\,\%)} 
        & 1940  & 97  & 1852  & 1883  & 1903 \\
\hline
\end{tabular}
\end{center}
\end{table}

\begin{figure}[htb]
\includegraphics[bbllx=30,bblly=20,bburx=475,bbury=800,width=\columnwidth,clip=]{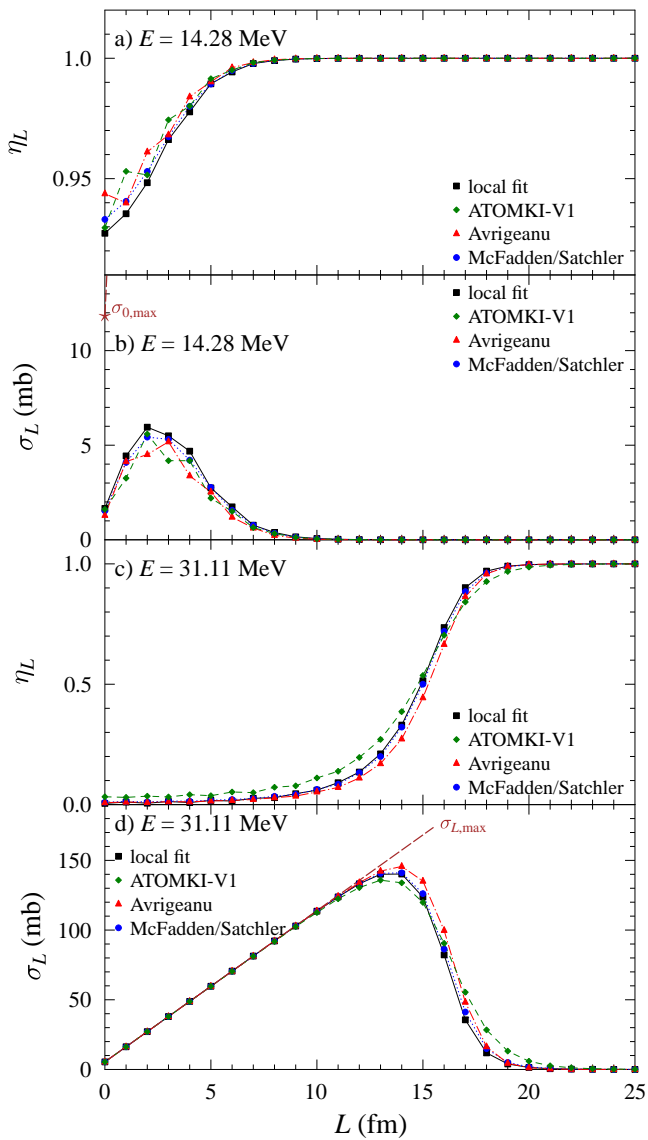}
\caption{
\label{fig:phase}
(Color online)
Reflexion coefficients $\eta_L$ (a,c) and contributions $\sigma_L$ of the
$L$-th partial wave to the total reaction cross section \stot\ (b,d) for the 
angular distributions at 14.28\,MeV (a,b) and 31.11\,MeV (c,d). The data are
obtained from the local fit (full black squares) and from
the global potentials under study \cite{McF66,Avr10,Mohr12}
(MCF: blue circles; AVR: red triangles; ATOMKI-V1: green diamonds; colors are
identical to Fig.~\ref{fig:scat}). The data points are connected by thin
colored lines to guide the eye. The maximum values $\sigma_{L,{\rm{max}}}$ are 
indicated as thick brown line (31.11\,MeV, d) and as data point for
$\sigma_{0,{\rm{max}}}$ (14.28\,MeV, b). Further discussion see text.
}
\end{figure}

The similarity of the total reaction cross sections \stot\ from the different
global potentials can be understood from a semi-classical interpretation of
the reflexion coefficients $\eta_L$ in the partial wave analysis. The
following discussion will focus on the shown energies of 31.11\,MeV above the
Coulomb barrier and 14.28\,MeV below the Coulomb barrier (see
Fig.~\ref{fig:pot} for the potentials and Fig.~\ref{fig:phase} for the
reflexion coefficients $\eta_L$ and the contributions $\sigma_L$ of the $L$-th
partial wave to \stot ). The discussion extends a similar
study in \cite{Mohr11}. 

Small angular momenta $L$ correspond to small impact parameters or central
collisions. At energies above the Coulomb barrier the incoming \al\ particles
are able to reach the interior, and the corresponding partial waves are almost
fully absorbed: $\eta_L \approx 0$ for $L \lesssim 10$ and thus $\sigma_L
\approx \sigma_{L,{\rm{max}}} = (2L+1) \pi/k^2$. The linear increase of
$\sigma_L$ is nicely seen in Fig.~\ref{fig:phase} (indicated by the brown
line). Consequently, only very few \al\ particles are scattered into the
backward angular region where the cross section is found to be much smaller
than the Rutherford cross section.

Large angular momenta $L$ correspond to large impact parameters or peripheral
collisions and scattering to forward angles. These partial waves do not reach
the nuclear interior and are thus practically not absorbed: $\eta_L \approx 1$
and $\sigma_L \approx 0$ is found for $L \gtrsim 20$.

Potentials have to fulfill two simple criteria to provide \stot\ above the
Coulomb 
barrier with sufficient accuracy. First, the potential must have the correct
short range; then $\eta_L \approx 1$ for $L \gtrsim 20$ is automatically
obtained. Second, the imaginary part must be sufficiently strong to guarantee
$\eta_L \approx 0$ for $L \lesssim 10$. These two simple criteria are
necessary but not yet sufficient for an excellent \al -nucleus potential which
should simultaneously describe elastic angular distributions with reasonable
$\chi^2/F$. Both simple criteria are fulfilled by
all global potentials in this study. The origin for differences in \stot\ from
different realistic potentials can only stem from a few $\eta_L$ or $\sigma_L$
for intermediate angular momenta $10 \lesssim L \lesssim 20$. But even in this
intermediate $L$ range the $\sigma_L$ have to decrease from
$\sigma_{L,{\rm{max}}}$ down to $\sigma_L \approx 0$. Thus, it is not at all
surprising that different global potentials provide very similar \stot\ at
energies above the Coulomb barrier.
As a test I have increased the imaginary part of the local potential at
31.11\,MeV by a factor of 2 and a factor of 10 (!). As expected, this leads to
only very moderate enhancements of the total reaction cross section \stot\ by
about 1\,\% and 8\,\% although the corresponding elastic angular distributions
change by more than one order of magnitude in the backward angular region.

The situation is much different at lower energies below the Coulomb
barrier. Again, partial waves with large angular momenta $L$ are not much
affected by any short-range potential, and $\eta_L \approx 1$ is
found. However, even 
\al\ particles with small impact parameters will mostly not be able to tunnel
through the Coulomb barrier. Thus, the elastic cross section at backward
angles rises and approaches the Rutherford cross section, and the $\eta_L$
remain close to unity even for small $L$. The total cross section \stot\ is
now composed of the contributions from a few partial waves with low $L
\lesssim 8$ in the chosen example of 14.28\,MeV. It is surprising that even
under these conditions the various potentials under study provide very similar
\stot\ (see Table \ref{tab:sigtot}), and indeed significant variations have
been found in the analysis of the recent \pras \ran \pmiv\ reaction data at
slightly lower energies \cite{Sau11,Mohr11}.

\subsection{Sensitivity of \stot\ to details of the optical potential}
\label{sec:sens}
Let me now artificially decompose the absorption into an interior contribution
($r \lesssim R_\alpha + R_{^{140}{\rm{Ce}}}$) and an exterior contribution ($r
\gg R_\alpha + R_{^{140}{\rm{Ce}}}$). It is the aim of this analysis to find
out the radial range of the potential which mainly defines \stot . A
semi-classical decomposition as e.g.\ presented in \cite{Bri77} may be
misleading here because the semi-classical approximation deviates from a fully
quantum-mechanical calculation by several per cent \cite{Bri77} which is not
acceptable at very low energies. For completeness it has also to be pointed out
that ``exterior'' in the above sense means radii above $\approx (10 - 12)$\,fm
where the nuclear real and imaginary potentials have practically dropped to
values close to zero (see Fig.~\ref{fig:pot}). Consequently, ``interior''
corresponds to the radial range where the nuclear potentials deviate from
zero, and ``interior absorption'' is dominated by the surface absorption by
the surface imaginary potential.

The interior contribution requires that
the incoming \al -particle tunnels through the Coulomb barrier and reaches an
area where it is absorbed by the imaginary part of the potential. It is thus
mainly sensitive to the shape of the Coulomb barrier which in turn is defined
by the exterior of the real part of the nuclear potential whereas details of
the imaginary part are not so important. This is similar to the full
absorption for small $L$ at higher energies, but only very few \al\ particles
are able to tunnel through the Coulomb barrier at very low energies. An
exterior contribution is obtained from the tail of the imaginary potential for
large radii; it does practically not depend on the real part. This tail is weak 
but affects all incoming \al\ particles. The relative importance of the
interior and exterior contributions can be estimated by a variation of the
strengths of the real and imaginary nuclear potentials or by introducing a
cutoff radius $R_{\rm{cut}}$ for the imaginary potential ($W(r) = 0$ for $r >
R_{\rm{cut}}$). 

It is found that at
energies around 12\,MeV an enhancement of the real part by a factor of two
increases \stot\ by a factor of about 3 whereas an enhancement of the same
factor of two for the
imaginary part increases \stot\ only by about 40\,\%; i.e., the interior
absorption is dominating at these energies. This is confirmed by the fact that
a cutoff for the imaginary potential at $R_{\rm{cut}} = 12$\,fm reduces
\stot\ by less than 10\,\%.

At much lower energies the
relative contribution of the exterior absorption increases, and at about
6\,MeV it is found that a factor of two enhancement of the real part and the
imaginary part both lead to the same factor of two enhancement for \stot . A
cutoff radius of $R_{\rm{cut}} = 12$\,fm for the imaginary part now reduces
\stot\ by about a factor of 6, and even a huge $R_{\rm{cut}} = 15$\,fm still
leads to a reduction of \stot\ by almost a factor of two.

In other words this means that the prediction of \stot\ at very low energies
requires detailed knowledge of the nuclear potential at very large
distances. The 
real part at large distances is essential for the internal contribution at
energies of about $10 - 15$\,MeV, and the imaginary part at large distances is
important for the external contribution which dominates at very low energies
far below 10\,MeV. However, the sensitivity of elastic
scattering data to the potential at large radii is limited. It is obvious that
the reproduction of \stot\ is a necessary requirement for any global
potential; but this requirement is almost automatically fulfilled at energies
above the Coulomb barrier. Further investigation is needed on the question
whether such a global potential is able to predict reaction cross sections of
\al -induced reactions below the Coulomb barrier. It remains an open question
which deviation (e.g.\ in $\chi^2/F$)
remains allowed in the description of elastic scattering data
to ensure a reasonable prediction of low-energy \al -induced reaction cross
sections. In this context an interesting example is the potential of
Fr\"ohlich and Rauscher \cite{Rau03}; it has been optimized for the
calculation of reaction cross sections and works very well for this purpose,
but it shows larger deviations than other global potentials in the
analysis of elastic scattering angular distributions.

Table \ref{tab:sigtot} shows that the relative uncertainty of \stot\ increases
strongly 
with decreasing energy. This is not surprising because the elastic scattering
angular distribution approaches the Rutherford cross section at low
energies. This holds in particular for the lowest energy where the ratio to
the Rutherford cross section is above 0.8 even at the most backward angles;
thus, very precise data are required here. Unfortunately, these data of Watson
{\it{et al.}}\ \cite{Wat71} are not available numerically and have been
extracted from Fig.~2 of \cite{Wat71} in the EXFOR \cite{EXFOR}
database. Using these EXFOR data leads to poor fits 
with parameters outside the expected range and \stot\ $\approx 65$\,mb. I have
carefully repeated the digitization of the data in Fig.~2 of \cite{Wat71}, and
I have 
obtained significantly larger elastic cross sections at backward angles. Now the
fit provides parameters in the expected range, and I find
a smaller total reaction cross section of \stot\ = 28.1\,mb. It
is interesting to note that the original analysis in \cite{Wat71} using
a Woods-Saxon potential gives an almost identical total reaction
cross section of \stot\ = 27.8\,mb. This confirms the newly digitized
data. These revised data will be sent to EXFOR. 

\begin{figure}
\includegraphics[width=\columnwidth,clip=]{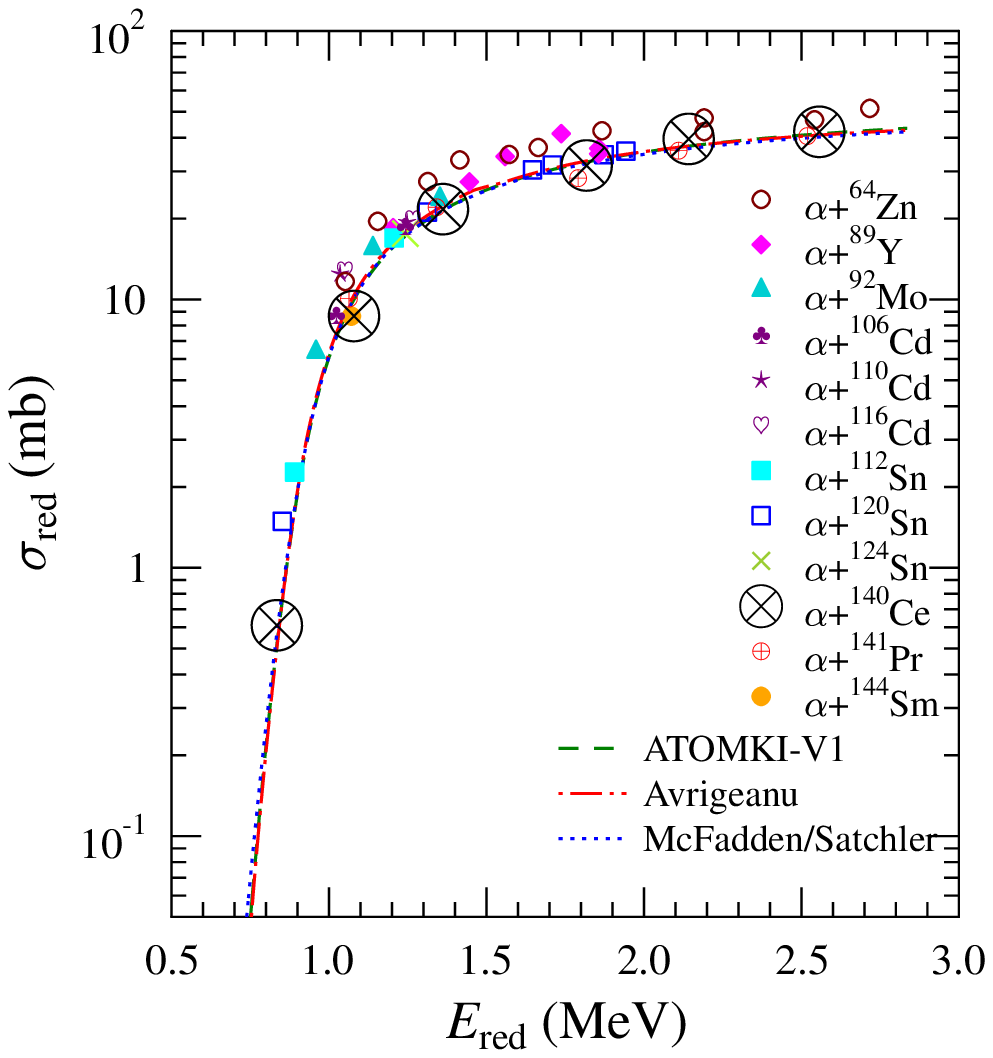}
\caption{
\label{fig:sigred}
(Color online)
Reduced reaction cross sections $\sigma_{\rm{red}}$ versus reduced energy
$E_{\rm{red}}$ for various \al -nucleus systems \cite{Mohr10,Mohr11,Mohr12},
compared to predictions from global \al -nucleus potentials
\cite{McF66,Avr10,Mohr12}. The experimental data are taken from
\cite{Wat71,Gua94} for \cer\ and from \cite{Bak72} for natural cerium.
The predictions are almost identical in the shown energy range. The indirect
data point at 10\,MeV from the $^{143}$Nd\rna \cer\ reaction is located far
below the shown region ($E_{\rm{red}} = 0.584$\,MeV, $\sigma_{\rm{red}} = 4.0
\times 10^{-5}$\,mb).
}
\end{figure}

For comparison of various targets at different energies, often reduced total
reaction cross sections \sred\ = \stot $/(A_P^{1/3}+A_T^{1/3})^2$ are plotted
versus the reduced energy \ered\ = $(A_P^{1/3}+A_T^{1/3}) E_{\rm{c.m.}}/(Z_P
Z_T)$, see e.g.\ \cite{Far10,Mohr10}. The data for many \al -nucleus systems
are shown in Fig.~\ref{fig:sigred}. It is obvious that the new data for
\cer\ fit nicely into the global systematics. Furthermore, it can be seen that
various global potentials \cite{McF66,Avr10,Mohr12} provide almost
identical \sred\ in the energy range above \ered\ $\gtrsim 0.8$\,MeV
($E_{\rm{c.m.}} \gtrsim 13.7$\,MeV) whereas
significant deviations between the different potentials are only found at very
low energies. It should be kept in mind that at these very low energies 
(\ered\ $\ll 0.8$\,MeV) the total reaction cross section is dominated by
inelastic scattering, mainly by Coulomb excitation, but not by compound
formation. 

\subsection{Lower energy limit for the determination of \stot }
\label{sec:low}
The extraction of \stot\ from elastic scattering angular distributions is
possible with reasonable uncertainties as long as the deviation from the
Rutherford cross section exceeds the experimental uncertainties
significantly. The Watson 
data at 15\,MeV fulfill this requirement with $\sigma/\sigma_R \approx 0.8$
(i.e., a deviation from Rutherford of about 20\,\%) at backward angles and
claimed experimental uncertainties of less than 5\,\%. However, at slightly
lower energies the extraction of \stot\ becomes impossible even with typical
uncertainties of up-to-date high-precision scattering data \cite{Mohr12,Pal12}
which are still of the order of a few per cent (systematic plus
statistical uncertainties).

\begin{figure}[htb]
\includegraphics[bbllx=25,bblly=10,bburx=568,bbury=782,width=\columnwidth,clip=]{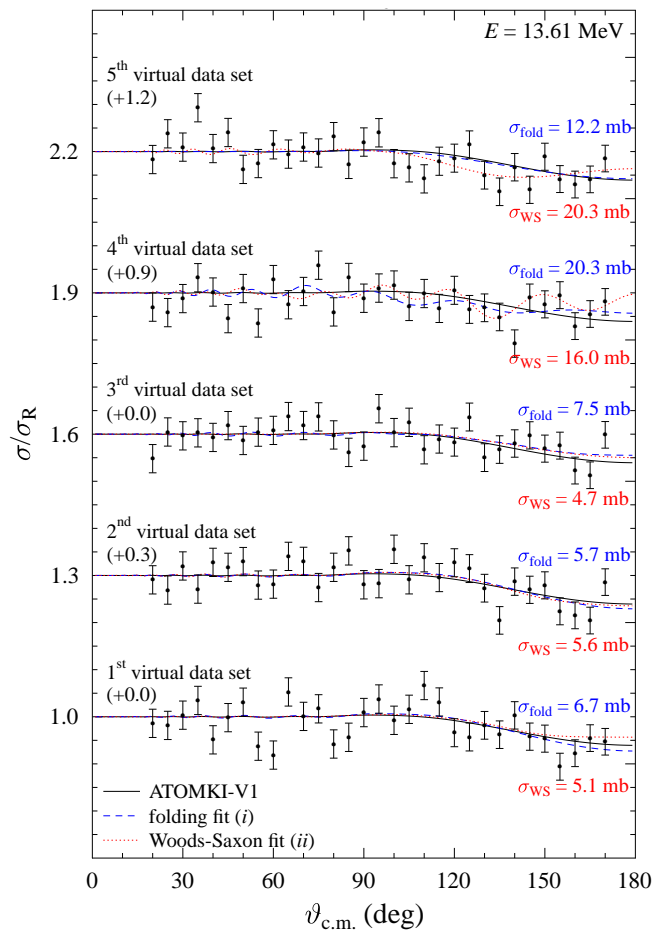}
\caption{
\label{fig:virt}
(Color online)
Rutherford normalized elastic scattering cross sections of \cer \raa
\cer\ versus $\vartheta_{\rm{c.m.}}$ at $E_{\rm{c.m.}} =
13.61$\,MeV. Various virtual data sets are created from the ATOMKI-V1
potential (full black line) by a random variation of the data points within an
assumed 3\,\% uncertainty, and they are analyzed to estimate the
uncertainty of the derived total reaction cross section \stot\ using either a
folding potential in the real part (case $i$, blue dashed) or volume
Woods-Saxon potentials (case $ii$, red dotted). Further
discussion see text. 
}
\end{figure}

A test calculation has been made at $E_{\rm{lab}} = 14.0$\,MeV ($E_{\rm{c.m.}} =
13.61$\,MeV) in the following way. First, the angular distribution at this
energy is calculated from the ATOMKI-V1 potential from $20^\circ$ to
$175^\circ$ in steps of $5^\circ$. Here I find
$\sigma/\sigma_R \approx 0.94$ at the most backward angles.
Next, several virtual
experimental data sets are created from this angular distribution by randomly
varying each data point within its uncertainty (Gaussian distribution with an
assumed 3\,\% 1-$\sigma$ uncertainty, corresponding to the most precise
available scattering data \cite{Mohr12,Pal12}). 
Finally, the various virtual data sets are
analyzed in the usual way by fitting the parameters of the optical
potential. Here I use two parametrizations: ($i$) a folding potential in the
real part and a surface Woods-Saxon potential in the imaginary part (i.e., the
same parametrization as the underlying ATOMKI-V1 potential), and
($ii$) a volume Woods-Saxon potential in the real and in the imaginary
part. The results are shown in Fig.~\ref{fig:virt}.
As expected, it is found that the fits become relatively unstable,
and restrictions on the number of fitted parameters have to be used. The
obtained total reaction cross sections from the fits to the virtual data sets
vary between about 5 and 20\,mb
(i.e.\ vary by a factor of four) whereas the starting value from the ATOMKI-V1
potential was 8.8\,mb. Thus, the uncertainty for the derived \stot\ is at
least a factor of two at 
the low energy of $E_{\rm{c.m.}} = 13.61$\,MeV which is dramatically larger
than the about 15\,\% uncertainty obtained at the slightly higher energy of
14.28\,MeV. However, the largest deviations for \stot\ are correlated with
extreme fitting parameters and unusually oscillating angular distributions
(e.g.\ in the case of the 4$^{\rm{th}}$ virtual data set). Constraining the
resulting parameters to a reasonable range may allow to determine \stot\ even in
the case of the 4$^{\rm{th}}$ virtual data set; but such constraints put into
question whether this is still a model-independent determination of \stot . 

A simple constraint might be to fix the geometry of the potentials from data
at higher energies, e.g.\ by taking the average values of $w$ in the real part
and $R_S$ and $a_S$ in the imaginary part from Table \ref{tab:pot}. Then only
two parameters ($\lambda$ and $W_S$) remain to be adjusted to the data
sets. However, even in this very restricted parameter space total reaction
cross sections \stot\ between about 5 and 20\,mb have been found, and it
remains necessary to keep $\lambda$ and $W_S$ close to their expected values
to obtain values of \stot\ close to the initial \stot\ = 8.8\,mb. So the lower
limit for the determination of \stot\ seems to be reached around $E \approx
13.5$\,MeV even if high-quality scattering data are available over the full
angular range.

\section{A short note on the $^{143}$Nd(${\rm{n}}$,\al )\cer\ reaction}
\label{sec:na}
\begin{figure}[htb]
\includegraphics[bbllx=35,bblly=15,bburx=475,bbury=310,width=\columnwidth,clip=]{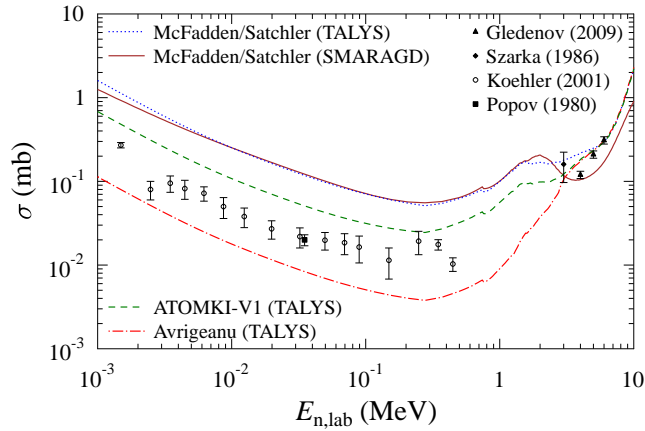}
\caption{
\label{fig:na}
(Color online)
Cross section of the $^{143}$Nd\rna \cer\ reaction. The experimental data are
taken from \cite{Gle09,Sza86,Koe01,Pop80}. Calculations have been performed
using the MCF potential in the statistical model codes {\sc{Talys}}
\cite{TALYS} (blue dotted) and {\sc{Smaragd}} \cite{SMARAGD} (full brown). The
results for the ATOMKI-V1 (green dashed) and AVR (red dash-dotted) potentials
are derived from the {\sc{Talys}} results. Further discussion see text.
}
\end{figure}

Recently, experimental data for the $^{143}$Nd\rna \cer\ reaction have been
measured at energies below 1\,MeV \cite{Koe01} and between 4 and 6\,MeV
\cite{Gle09}, and earlier data below 10\,MeV are available from \cite{Sza86}
and \cite{Pop80}.
The calculation of the $^{143}$Nd\rna \cer\ reaction cross section in the
statistical model requires the 
knowledge of the \cer -\al\ potential. Because of the huge positive $Q$-value
of 9.72\,MeV, this potential has to be determined at energies around 15\,MeV
for the higher-energy data in \cite{Gle09}. This is exactly the energy range
of the Watson {\it et al.}\ scattering data \cite{Wat71} where the various
potentials under study were able to reproduce \stot\ within about $15-20$\,\%
(see Table \ref{tab:sigtot}). Thus, the uncertainty from the \al -nucleus
potential in 
the calculation of the $^{143}$Nd\rna \cer\ reaction remains very limited. I
have calculated the cross sections using the default
parameters of {\sc{Talys V1.4}} \cite{TALYS} and the MCF potential in the
\al\ channel. The result is shown in Fig.~\ref{fig:na}. The agreement with the
experimental data above 1\,MeV \cite{Gle09,Sza86} is reasonable although the
energy dependence of the data points is not perfectly reproduced. At energies
below 1\,MeV the experimental data of \cite{Koe01,Pop80} are overestimated
which is a typical behavior for the MCF potential at energies significantly
below the Coulomb barrier, see e.g.\ for $^{144}$Sm\rag $^{148}$Gd
\cite{Som98}, $^{141}$Pr\ran $^{144}$Pm \cite{Sau11,Mohr11}, $^{112}$Sn\rag
$^{116}$Te \cite{Ozk07}, and $^{106}$Cd\rag $^{110}$Cd \cite{Gyu06}.

The calculations have been repeated using the code {\sc{Smaragd}}
\cite{SMARAGD} with its default parameters (here the MCF potential is used in
the \al\ channel by default), see Fig.~\ref{fig:na}, brown line. As pointed
out in \cite{Rau12}, the 
$^{143}$Nd\rna \cer\ reaction cross section is mainly sensitive to the optical
potential in the \al\ channel whereas the other ingredients of the statistical
model codes have only minor influence on the \rna\ cross section (for details
see the sensitivity figures shown in Ref.~\cite{RauWEB}). 
As expected, the two codes provide very similar results
for energies below 1\,MeV. However, above 1\,MeV surprisingly significant
differences up to a factor of two appear. Interestingly, the {\sc{Smaragd}}
calculation shows a similar energy dependence as the experimental data points
\cite{Gle09,Sza86} but it underestimates the abolute scale significantly.
It is beyond the scope of the
present paper to provide a detailed comparison of the {\sc{Talys}} and
{\sc{Smaragd}} codes. The disagreement between both codes may be related to the
treatment of excited states in the exit channel (i.e., in the \cer\ nucleus);
this sensitivity is not explicitly studied in \cite{RauWEB}. But different
choices of this treatment in {\sc{Talys}} lead only to minor variations of
about 20\,\% for the \rna\ cross section and cannot explain the factor of two
discrepancy between {\sc{Talys}} and {\sc{Smaragd}}. Therefore, great care is
required in the determination of the \al -nucleus potential from low-energy
reaction data because the influence of other ingredients on the calculated
cross sections has to be investigated in detail.

Nevertheless, the influence of different potentials on the $^{143}$Nd\rna
\cer\ reaction cross section can be studied at low energies below 1\,MeV where
{\sc{Talys}} and {\sc{Smaragd}} show the expected agreement for the MCF
potential. It is found that the MCF potential overestimates the experimental
data by a factor of 5. As this energy range below 1\,MeV corresponds to
energies in the \al -channel of about 10\,MeV, it is possible to estimate
\stot\ at this energy because \stot\ is only sensitive to the chosen \al
-nucleus potential: \stot\ $\approx 1.9 \pm 0.4\,\mu$b. For comparison, the 
predictions from the different potentials under study at 10\,MeV are also
listed in Table \ref{tab:sigtot}. 

It is technically difficult to use the ATOMKI-V1 folding potential in the
{\sc{Talys}} or {\sc{Smaragd}} codes. However, because the \rna\ cross section
is practically sensitive only to the \al -potential \cite{Rau12,RauWEB}, it is
possible to calculate the $^{143}$Nd\rna \cer\ cross section for the ATOMKI-V1
potential by scaling the MCF result with the ratio of the total cross sections
\stot\ of the ATOMKI-V1 potential and the MCF potential at the corresponding
energy in the \al\ channel. The same method has been applied for the AVR
potential. In both cases the underlying MCF result has been taken from the
{\sc{Talys}} calculation. The results are compared to the experimental data of
\cite{Gle09,Koe01,Sza86,Pop80} in Fig.~\ref{fig:na}. At energies below 1\,MeV
the ATOMKI-V1 potential slightly overestimates the experimental data of
\cite{Koe01,Pop80} whereas the AVR potential slightly underestimates the
data. But both potentials are much closer to the experimental data than the MCF
potential which predicts the \rna\ cross section about a factor of two higher
than the ATOMKI-V1 potential and more than one order of magnitude higher then
the AVR potential.

For the data in the MeV region \cite{Gle09,Sza86} (corresponding
to the 15\,MeV region in the \al\ channel) all potentials
under study provide almost identical total reaction cross sections \stot , and
thus the calculated \rna\ cross sections are very similar. It is not
possible to extract information on the potential in this energy range. This
holds in particular as long as significant differences appear between the two
widely used codes {\sc{Talys}} and {\sc{Smaragd}}.
Finally, it is interesting to note that almost the same behavior of the MCF
(overestimation), 
AVR (slight underestimation), and ATOMKI-V1 (slight overestimation; not shown
in \cite{Mohr11}) potentials is found in the analysis \cite{Mohr11} of the
recent \pras \ran \pmiv\ data \cite{Sau11}.

\section{Summary and conclusions}
\label{sec:summ}
Angular distributions of \cer \raa \cer\ elastic scattering
were analyzed in the framework of the optical model. Excellent fits were
obtained using locally adjusted parameters of the real folding and imaginary
Woods-Saxon potentials. Total reaction cross sections \stot\ were derived from
the local fits using Eq.~(\ref{eq:stot}). The new data are in excellent
agreement with the systematics of reduced cross sections
\sred\ \cite{Mohr10,Mohr12}.

It was shown that various global \al
-nucleus potentials predict the total reaction cross sections \stot\ very well
although the predicted angular distributions do not agree perfectly with the
experimental angular distributions and may even deviate by up to an order of
magnitude at very backward angles. Thus, at least at energies above about
15\,MeV, the \al -nucleus potential is not the major source of uncertainties
in the calculation of \al -induced reaction cross sections in the statistical
model. This can be understood by a semi-classical interpretation of the
partial wave analysis. 

It is highlighted that data with very
small uncertainties are required for the extraction of \stot\ at low energies
where the elastic scattering angular distribution approaches the Rutherford
cross section. As the analysis of the Watson {\it et al.}\ data \cite{Wat71}
shows, special diligence is required for data which are taken from
published figures by digitization.
An extraction of \stot\ from elastic scattering angular distributions 
becomes impossible with a reasonable uncertainty as soon as the deviation from
the Rutherford cross section at backward angles becomes smaller than the
experimental uncertainty.

Finally, further information on the \al -nucleus potentials at very low
energies can be 
extracted only from the analysis of reaction data. Although great care is
required for such an analysis because of the influence of other ingredients of
the statistical model calculations, it can be concluded from the low-energy
$^{143}$Nd\rna \cer\ data below $E_{\rm{n}} = 1$\,MeV that the recent ATOMKI-V1
and AVR potentials provide a significant improvement around $E \approx
10$\,MeV in the \al\ channel, i.e.\ far below the Coulomb barrier, whereas the
MCF potential overestimates the $^{143}$Nd\rna \cer\ data by a factor of 5.

\begin{acknowledgments}
I thank P.\ Guazzoni and L.\ Zetta for providing their experimental \cer \raa
\cer\ scattering data before publication, R.\ Lichtenth\"aler for
providing the phase shift fitting code {\sc{Para}} of \cite{Chi96}, and
T.\ Rauscher for the {\sc{Smaragd}} calculations.
This work was supported by OTKA (NN83261).
\end{acknowledgments}


\vspace{5cm}
\begin{thebibliography}{99}
%
\bibitem{Som98}
  E.\ Somorjai, Zs.\ F\"ul\"op, A.\ Z.\ Kiss, C.\ E.\ Rolfs,
  H.-P.\ Trautvetter, U.\ Greife, M.\ Junker, S.\ Goriely, M.\ Arnould,
  M.\ Rayet, T.\ Rauscher, H.\ Oberhummer,
  Astron.\ Astrophys.\ {\bf 333}, 1112 (1998).
%
\bibitem{Gyu06} 
Gy.\,Gy\"urky, G.\ G.\ Kiss, Z.\ Elekes, Zs. F\"ul\"op, E.\ Somorjai,
A.\ Palumbo, J.\ G\"orres, H.\ Y.\ Lee, W.\ Rapp, M.\ Wiescher, N.\ \"Ozkan,
R.\ T.\ G\"uray, G.\ Efe, T.\ Rauscher,
\prc\ {\bf 74}, 025805 (2006).
%
\bibitem{Ozk07}
  N.\ {\"O}zkan, G.\ Efe, R.\ T.\ G{\"u}ray, A.\ Palumbo, J.\ G{\"o}rres,
  H.\ Y.\ Lee, L.\ O.\ Lamm, W.\ Rapp, E.\ Stech, M.\ Wiescher,
  Gy.\ Gy{\"u}rky, Zs.\ F{\"u}l{\"o}p, E.\ Somorjai,
  \prc\ {\bf 75}, 025801 (2007).
%
\bibitem{Cat08} 
I.\ Cata-Danil, D.\ Filipescu, M.\ Ivascu, D.\ Bucurescu, N.\ V.\ Zamfir,
T.\ Glodariu, L.\ Stroe, G.\ Cata-Danil, D.\ G.\ Ghita, C.\ Mihai,
G.\ Suliman, T.\ Sava, 
\prc\ {\bf 78}, 035803 (2008).
%
\bibitem{Yal09} 
C.\ Yalcin,  R.\ T.\ G\"uray, N.\ \"Ozkan, S.\ Kutlu, Gy.\ Gy\"urky,
J.\ Farkas, G.\ G.\ Kiss, Zs.\ F\"ul\"op, A.\ Simon, E.\ Somorjai,
T.\ Rauscher,
\prc\ {\bf 79}, 065801 (2009).
%
\bibitem{Gyu10} 
Gy.\ Gy\"urky, Z.\ Elekes, J.\ Farkas, Zs.\ F\"ul\"op, Z.\ Hal\'asz,
G.\ G.\ Kiss, E.\ Somorjai, T.\ Sz\"ucs, R.\ T.\ G\"uray, N.\ \"Ozkan,
C.\ Yalcin and T.\ Rauscher, 
J.\ Phys.\ G {\bf 37}, 115201 (2010).
%
\bibitem{Kis11}
G.\ G.\ Kiss, T.\ Rauscher, T.\ Sz\"ucs, Zs.\ Kert\'esz, Zs.\ F\"ul\"op,
Gy.\ Gy\"urky, C.\ Fr\"ohlich, J.\ Farkas, Z.\ Elekes, E.\ Somorjai, 
Phys.\ Lett.\ B {\bf 695}, 419 (2011).
%
\bibitem{Kis12}
G.\ G.\ Kiss, T.\ Sz\"ucs, Zs.\ T\"or\"ok, Z.\ Korkulu, 
Gy.\ Gy\"urky, Z.\ Hal\'asz, Zs.\ F\"ul\"op, E.\ Somorjai, T.\ Rauscher, 
\prc\ {\bf 86}, 035801 (2012).
%
\bibitem{Sau11}
A.\ Sauerwein, H.\ W.\ Becker, H.\ Dombrowski, M.\ Elvers, J.\ Endres,
U.\ Giesen, J.\ Hasper, A.\ Hennig, L.\ Netterdon, T.\ Rauscher, D.\ Rogalla,
K.\ O.\ Zell, A.\ Zilges,
\prc\ {\bf 84}, 045808 (2011).
%
\bibitem{Mohr11}
P.\ Mohr,
\prc\ {\bf 84}, 055803 (2011).
%
\bibitem{Gyu12}
Gy.\ Gy\"urky, P.\ Mohr, Zs.\ F\"ul\"op, Z.\ Hal\'asz, G.\ G.\ Kiss,
T.\ Sz\"ucs, E.\ Somorjai,
\prc\ {\bf 86}, 041601(R) (2012).
%
\bibitem{Wat71}
B.\ D.\ Watson, D.\ Robson, D.\ D.\ Tolbert, R.\ H.\ Davis,
\prc\ {\bf 4}, 2240 (1971). 
%
\bibitem{Gua94}
P.\ Guazzoni and L.\ Zetta,
{\it{private communication to G.\ Staudt}}, 1994 (unpublished).
%
\bibitem{EXFOR}
EXFOR data base, 
{\it{http://www-nds.iaea.org/exfor/exfor.htm}}, 
Version September 21, 2012.
%
\bibitem{Bak72}
F.\ T.\ Baker and R.\ Tickle,
\prc\ {\bf 5}, 182 (1972).
%
\bibitem{Avr10}
M.\ Avrigeanu and V.\ Avrigeanu,
\prc\ {\bf 82}, 014606 (2010).
%
\bibitem{Avr09}
M.\ Avrigeanu, A.\ C.\ Obreja, F.\ L.\ Roman, V.\ Avrigeanu, W.\ von Oertzen,
At.\ Data Nucl.\ Data Tables {\bf 95}, 501 (2009).
%
\bibitem{Mohr12}
P.\ Mohr, G.\ G.\ Kiss, Zs.\ F\"ul\"op, D.\ Galaviz, Gy.\ Gy\"urky, 
E.\ Somorjai,
At.\ Data Nucl.\ Data Tables, accepted for publication;
{\it{arXiv:1212.2891}}.
%
\bibitem{McF66}
L.\ McFadden and G.\ R.\ Satchler, 
Nucl.\ Phys.\ {\bf 84}, 177 (1966).
%
\bibitem{Atz96} 
  U.\ Atzrott, P.\ Mohr, H.\ Abele, C.\ Hillenmayer, and
  G.\ Staudt,
  \prc\ {\bf 53}, 1336 (1996).
%
\bibitem{Vri87}
H.\ de Vries, C.\ W.\ de Jager, and C.\ de Vries,
At.\ Data and Nucl.\ Data Tables {\bf 36}, 495 (1987).
%
\bibitem{Mil88}
B.\ L.\ Miller, L.\ S.\ Cardman, C.\ N.\ Papanicolas,
T.\ E.\ Milliman, J.\ P.\ Connelly, J.\ H.\ Heisenberg, F.\ W.\ Hersman,
J.\ E.\ Wise, B.\ Frois, D.\ Goutte, V.\ Meot,
\prc\ {\bf 37}, 895 (1988).
%
\bibitem{Kim92}
W.\ Kim, B.\ L.\ Miller, J.\ R.\ Calarco, L.\ S.\ Cardman, J.\ P.\ Connelly,
S.\ A.\ Fayans, B.\ Frois, D.\ Goutte, J.\ H.\ Heisenberg, F.\ W.\ Hersman,
V.\ Meot, T.\ E.\ Milliman, P.\ Mueller, C.\ N.\ Papanicolas,
A.\ P.\ Platonov, V.\ Yu.\ Ponomarev, J.\ E.\ Wise,
\prc\ {\bf 45}, 2290 (1992).
%
\bibitem{Mil94}
  B.\ L.\ Miller, 
  Ph.D.\ thesis,
  University of Illinois at Urbana-Champaign, 1994 (unpublished);
  available at {\it{http://hdl.handle.net/2142/18865}}.
%
\bibitem{Mohr00}
P.\ Mohr,
\prc\ {\bf 61}, 045802 (2000).
%
\bibitem{Chi96}
V.\ Chist{\'e}, R.\ Lichtenth{\"a}ler, A.\ C.\ C.\ Villari, L.\ C.\ Gomes,
\prc\ {\bf 54}, 784 (1996).
%
\bibitem{Mohr10}
P.\,Mohr, D.\,Galaviz, Zs.\,F\"ul\"op, Gy.\,Gy\"urky, G.\,G.\,Kiss,
E.\,Somorjai,
\prc\ {\bf 82}, {047601} (2010).
%
\bibitem{Bri77}
D.\ M.\ Brink and N.\ Takigawa,
Nucl.\ Phys.\ {\bf A279}, 159 (1977).
%

\bibitem{Far10}
P.\ N.\ de Faria {\it et al.},
\prc {\bf 81}, 044605 (2010).  
%
\bibitem{Pal12}
A.\ Palumbo {\it et al.},
\prc\ {\bf 85}, 035808 (2012).
%
\bibitem{Rau03}
T.\ Rauscher,
Nucl.\ Phys.\ {\bf A719}, 73c (2003).
%
\bibitem{Koe01}
P.\ E.\ Koehler, Yu.\ M.\ Gledenov, J.\ Andrzejweski, K.\ H.\ Guber,
S.\ Raman, T.\ Rauscher,
Nucl.\ Phys.\ {\bf A688}, 86c (2001).
%
\bibitem{Gle09}
Yu.\ M.\ Gledenov, M.\ V.\ Sedysheva, V.\ A.\ Stolupin, Guohui Zhang, Jiaguo
Zhang, Hao Wu, Jiaming Liu, Jinxiang Chen, G.\ Khuukhenkhuu, P.\ E.\ Koehler,
P.\ J.\ Szalanski,
\prc\ {\bf 80}, 044602 (2009).
%
\bibitem{Sza86}
I.\ Szarka, M.\ Florek, J.\ Oravec, K.\ Hol{\'y},
Nucl.\ Inst.\ Meth.\ Phys.\ Res.\ B {\bf 17}, 472 (1986).
%
\bibitem{Pop80}
Yu.\ P.\ Popov, V.\ I.\ Salackij, G.\ Khuukhenkhuu,
Sov.\ J.\ Nucl.\ Phys.\ {\bf 32}, 459 (1980).
%
\bibitem{TALYS}
A.\ Koning, S.\ Hilaire, M.\ Duijvestijn, 
computer code {\sc{Talys}}, version 1.4, 
available online at {\it{http://www.talys.eu/}},
Proc.\ {\it{International Conference on Nuclear Data for Science and
    Technology}}, April 22-27, Nice, France, Ed.~O.\ Bersillon, F.\ Gunsing,
E.\ Bauge, R.\ Jacmin, S.\ Leray,
EDP Sciences, 2008, p.211-214.
%
\bibitem{SMARAGD}
T.\ Rauscher,
computer code {\sc{SMARAGD}},
{\it{http://www.nucastro.org/smaragd.html}}.
%
\bibitem{Rau12}
T.\ Rauscher,
Astroph.\ J.\ Suppl.\ {\bf 201}, 26 (2012).
%
\bibitem{RauWEB}
T.\ Rauscher,
{\it{http://www.nucastro.org}}.
%
\end{thebibliography}
\end{document}